\documentclass[twocolumn,aps,prc,superscriptaddress,reprint]{revtex4-1}
\makeatletter

\newcommand{\Rmnum}[1]{\expandafter\@slowromancap\romannumeral #1@}
\makeatother
\usepackage{blindtext}
\usepackage{graphicx}
\usepackage{dcolumn}
\usepackage{bm}
\usepackage{booktabs}
\usepackage{mathrsfs}
\usepackage{ulem}
\usepackage{verbatim}
\usepackage{indentfirst}
\usepackage{float}
\usepackage{subfigure}
\usepackage[colorlinks,
           bookmarks=true,
           linkcolor=blue,
           urlcolor=blue, 
            anchorcolor=black,
            citecolor=blue
            ]{hyperref}
\usepackage{hypcap}
\usepackage{amsmath}
\usepackage{lipsum}

\usepackage{epsfig}

\usepackage{amsmath}
\usepackage{float}
\usepackage{enumitem}   
\usepackage[colorlinks]{hyperref}
\newcommand{\bef}{\begin{figure}}
\newcommand{\eef}{\end{figure}}

\newcommand{\be}{\begin{equation}}
\newcommand{\ee}{\end{equation}}
\newcommand{\bea}{\begin{eqnarray}}
\newcommand{\eea}{\end{eqnarray}}
\newcommand{\vv}{v_{\rm 1}}
\newcommand{\sq}{\sqrt{s_{NN}}}
\newcommand{\dv}{dv_{\rm 1}/dy}

\begin{document}

\title{Energy dependence study of directed flow in Au+Au collisions using an improved coalescence in AMPT model}

\author{Kishora Nayak}
\affiliation{Key Laboratory of Quark \& Lepton Physics (MOE) and Institute of Particle Physics,\\
Central China Normal University, Wuhan 430079, China}
\author{Shusu Shi}
\email{shiss@mail.ccnu.edu.cn}
\affiliation{Key Laboratory of Quark \& Lepton Physics (MOE) and Institute of Particle Physics,\\
Central China Normal University, Wuhan 430079, China}
\author{Nu Xu}
\affiliation{Key Laboratory of Quark \& Lepton Physics (MOE) and Institute of Particle Physics,\\
Central China Normal University, Wuhan 430079, China}
\affiliation{Institute of Modern Physics,
Chinese Academy of Sciences, Lanzhou, China}
\author{Zi-Wei Lin}
\affiliation{Key Laboratory of Quark \& Lepton Physics (MOE) and Institute of Particle Physics,\\
Central China Normal University, Wuhan 430079, China}
\affiliation{Department of Physics, East Carolina University, Greenville, NC 27858, USA}


\date{\today}

\begin{abstract}
The rapidity-odd component of directed flow ($\vv$) of identified hardons ($\pi^{\pm}$, $K^{\pm}$, $K^{0}_{S}$, $p$, $\overline{p}$, $\phi$, $\Xi$, $\overline{\Xi}$, $\Lambda$, $\overline{\Lambda}$) and partons ($u$, $\overline{u}$, $d$, $\overline{d}$, $s$, $\overline{s}$) in Au+Au collisions at various beam energies ($\sq$ = 7.7, 11.5, 14.5, 19.6, 27, 39, 54.4, 62.4, 200 GeV) using a multi-phase transport model is analyzed. A data driven approach (inspired from the experimental analysis)  is performed here to distinguish the transported and produced quarks which are found to have different directed flow values at various collision beam energies. The coalescence sum rule (Number of Constituent Quark scaling) violation is observed at lower energies where hadronic matters dominate. The strange quark ($s$) and $\phi$ meson slope (d$\vv$/dy) show a double sign change around 14.5 GeV unlike other partons and hadrons. It suggests that strange quark is more sensitive to the softening of Equation of State (EoS).
      
\end{abstract}

\maketitle
\section{Introduction}
\label{Intro}
The main goal of relativistic heavy-ion collision experiments is to understand  the properties and evolution of strongly interacting matter, called the Quark-Gluon Plasma, as well as to explore the hadron-quark phase transition. The rapidity-odd component of directed flow ($\vv$) is an important probe to study the in-medium dynamics as it is sensitive to the equation of state (EoS) of the produced medium. Directed flow is generated during the nuclear passage time (2R/$\gamma$ $\sim$ 0.1 fm/$c$ at 200 GeV) and it probes the onset of bulk collective dynamics in the early stage of the collision~\cite{nuclPassage1,nuclPassage2}. As a suggested signature of a first order phase transition, directed flow is sensitive to the existance of the critical point and it plays an important role in the proposed beam energy scan program\cite{critTheo1,critTheo2,critTheo3,critTheo4,critTheo5,newAdd1}. The first-order harmonic of the Fourier expansion in momentum distribution of emitted particles is characterized as directed flow, 
\begin{equation}
\vv = \langle {\rm cos}(\phi-\Psi_{RP})\rangle
\label{eq1}
\end{equation}
where $\phi$ and $\Psi_{RP}$ are the azimuthal angle and reaction plane angle, respectively~\cite{flowDef1,flowDef2,flowDef3}. The $\vv$ contains both rapidity-odd and rapidity-even components. Rapidity-odd component ($v_{1}^{\it{odd}}(y)$ = -$v_{1}^{\it{odd}}(-y)$) is referred to the sideward collective motion of emitted hadrons with respect to collision reaction plane. The rapidity-even component even ($v_{1}^{\it{even}}(y)$ = $v_{1}^{even}(-y)$) is unrelated to the reaction plane and it originates from event-by-event fluctuations in the initial colliding nuclei. In this paper, $v_{1}(y)$ implicitly refers to the odd component of directed flow. The transport and hydrodynamic models calculations suggested that the directed flow of baryon $\vv$ at mid-rapidity (y$\sim$0) is sensitive to the equation of state of the system~\cite{critTheo6,critTheo2}. Severals hydrodynamic model calculations predict that the negative $\vv$-slope near mid-rapidity called as “wiggle” or “anti-flow” might be a possible QGP signature~\cite{wiggle1,wiggle2}. Number-of-constituent-quark (NCQ) scaling is an example of coalescence behavior among quarks. Because of the NCQ scaling, which is observed at RHIC~\cite{ncqSTAR, ncqPHENIX} and LHC~\cite{ncqALICE}, the higher order flow harmonics like $v_{2}$ behaves as if it is developed at the partonic  level~\cite{parton1,parton2,parton3}. There are recent experimental measurement of directed flow of various identified hadrons ($\pi^{\pm}$, $K^{\pm}$, $K^{0}_{S}$, p, $\bar{\rm p}$, $\phi$, $\Lambda$, $\bar{\Lambda}$) from the STAR collaboration at RHIC over a wide range of colliding beam energies (7.7-200 GeV)~\cite{starPRL}. Comprehensive $\vv$ measurement from STAR~\cite{starPRL} supports the coalescence mechanism as the dominant process in particle formation dynamics. There are several studies in heavy-ion collisions to understand the hadron and nuclei formation via coalescence and also hadronization of quarks in heavy-ion collisions~\cite{coal0,coal00,coal1,coal2,coal3,coal4,coal5,coal6}. In recent articles the importance of coalescence mechanism and energy dependence directed flow are discussed~\cite{dir1, dir2, dir3, dir4, dir5} and an experimental review of $\vv$ can be found in Ref.~\cite{review}.

The interplay between NCQ scaling and the transport of initial-state $u$ and $d$ quarks towards mid-rapidity during the collision offers possibilities for new insights~\cite{transport1}. The produced strange ($s$) and anti-strange ($\overline{s}$) quark contribute in the resonance ($\phi$) formation and hence also play vital role in understanding the particle formation mechanism. Understanding the strange quarks or particles are very important in order to understand the EoS, as the $dv_{1}/dy$ of $\phi$ meson also shows a hints of sign change similar to baryons ($p$, $\Lambda$)~\cite{starPRL}. An approach to study $\vv$ performed in this paper is inspired from the STAR experiment at RHIC~\cite{starPRL}, where a comprehensive measurement of directed flow of identified hadrons are reported in a range of collision energies. The experimental paper verified the coalescence sum rule (NCQ scaling) using $v_{1}$ measurement although the NCQ scaling is well known in elliptic flow ($v_{2}$) measurement of identified hadrons at RHIC and LHC~\cite{ncqSTAR,ncqPHENIX,ncqALICE,shi}. Our model calculation is also compared with the experimental results. The calculation can reasonably well describe the data for mesons over a range of energies. $\vv$ prediction for $\Xi$ and $\overline{\Xi}$ baryons are also given along with the new energy 54.4 GeV for various hadron species.  

The paper is organized in the following sections. Section II provides a brief description about the AMPT event generator~\cite{ampt}. The analysis details of calculating directed flow and the results which include the $\vv$ of partons and hadrons followed by the slope parameter (d$\vv$/dy) are discussed in the Sec. III. A summary with final remarks are given in the Sec. IV.  

\section{The AMPT model} \label{sec||}
A multi-phase transport model especially the string-melting version (AMPT-SM) is often used to understand the experimental heavy-ion collision results. The hot and dense matter formed due to relativistic heavy-ion collisions are expected to be in parton degrees of freedom and the AMPT-SM also evolves through the partonic medium, thus makes it a suitable model for interpreting the experimental results. The AMPT-SM version mainly consists of four parts. The initial conditions are taken from Heavy Ion Jet Inter-action Generator (HIJING)~\cite{hijing}. Scatterings among partons are described by Zhang’s parton cascade (ZPC)~\cite{zpc} model and for hadronisation it uses the coalescence model. An extended relativistic transport (ART) model  describes the final hadronic evolution~\cite{art}. HIJING model includes two body nucleon-nucleon interactions to form excited strings and mini jets via hard and soft processes. The mini-jet parton undergoes scattering before they fragment to partons and subsequently into hadrons. The partonic interaction in ZPC model is described by two body partonic elastic cross section ($\sigma_{p}$) as given in Eq.~\ref{eq2}.
\begin{equation}
\sigma_{p} = \frac{9\pi\alpha_{S}^{2}}{2\mu^{2}}
\label{eq2}
\end{equation}

In this study the strong coupling constant ($\alpha_{S}$) and parton screening mass ($\mu$) are set to be 0.33 and 3.20 fm$^{-1}$, respectively, leading to $\sigma_{p}$ = 1.5 mb. After partons freeze out, the hadronization process in AMPT is described by a quark coalescence model. A meson is formed by combining a quark with a nearby anti-quark. Similarly, three quarks (anti-quarks) combine to form a baryon (anti-baryon). Here the formation process of mesons or baryons (anti-baryons) is independent of the relative momentum among the coalescening partons. In this coalescence process, each number of baryons, anti-baryons and mesons in an event are conserved individually. However, in the present study an improved quark coalescence method has been used~\cite{newAMPT}. The constraint which forced separate conservation of the baryons, anti-baryons, and mesons number via the quark coalescence has been removed in the new method. However, the net-baryons and net-strangeness numbers are still conserved for each event. In the new coalescence model, for a meson formation, any available quark searches all available antiquarks and records the closest relative distance ($d_{M}$) as the potential coalescence partner. The quark also searches all available quarks to find the closest one in distance as a potential coalescence partner to form a baryon, and then searches all other available quarks again to find the one that gives the smallest average relative distance ($d_B$) among these three quarks. The condition $d_{B}$ $<$ $d_{M}$ * $r_{BM}$ has to be satisfied to form a baryon else a meson is formed. A new coalescence parameter  $r_{BM}$, controls the relative probability of quark to form a baryon rather than meson. The limit of $r_{BM} \rightarrow 0$ and $r_{BM} \rightarrow \infty$ corresponds to no anti-baryon formation (although to keep net-baryon number conservation, a minimum number of baryons would be formed) and almost no meson formation, respectively. Similar coalescence procedure is also applied to all anti-quarks. 

In this analysis the mean field is not included~\cite{meanfiled}. The new parameter $r_{BM}$ which controls the relative probability to form baryon via coalescence of a quark is set to 0.61. The popcorn parameter PARJ(5) value is changed to 0 from the default value 1.0 which controls the relative percentage of the $B\bar{B}$ and $BM\bar{M}$ channels. This $r_{BM}$ parameter value is able to describe the dN/dy of proton yields at mid-rapidity in central Au+Au collisions at $\sq$ = 200 GeV and central Pb+Pb collisions at $\sq$ = 2.76 TeV as shown in the Ref.~\cite{newAMPT}.  

\section{Analysis and Results} \label{sec||I} 
In this study, an improved version of AMPT-SM model $\sigma_{p}$ = 1.5 mb is used to study the directed flow of identified hadrons in mid-central (10-40$\%$) Au+Au collisions at $\sq$ = 7.7, 11.5, 14.5, 19.6, 27, 39, 54.4, 62.4, 200 GeV, corresponding to RHIC beam energy scan program (BES-I). The centrality is determined using the charged particle multiplicity ($|\eta|$ $<$ 0.5). This analysis is inspired from the recent experimental measurement from STAR at RHIC, where a comprehensive $\vv$ measurement has been performed and coalescence sum rule is verified. The effect of hadronic interaction on directed flow is also studied by changing the hadron cascade time ($t_{max}$) in the AMPT-SM. The particles reported here are identified from their PYTHIA-id (PID). The particle selection cuts (e.g. momentum p, transverse momentum $p_{T}$) are listed in the Tab. {\color{blue}{I}}, which is similar to the experimental data~\cite{protonv1, starPRL}, in order to have a better comparison.   

\begin{table}[h]
\begin{tabular}{lll|ll}
\cline{2-3}
\multicolumn{1}{l|}{} & \multicolumn{1}{l|}{Hadron} & \multicolumn{1}{l|}{$p_{T}$ cut (GeV/c)} &  &  \\ \cline{2-3}
\multicolumn{1}{l|}{} & \multicolumn{1}{l|}{$p$, $\bar{p}$} & \multicolumn{1}{l|}{$0.2<p_{T}<2.0$ } &  &  \\ \cline{2-3}
\multicolumn{1}{l|}{} & \multicolumn{1}{l|}{$\pi^{\pm}$, $K^{\pm}$} & \multicolumn{1}{l|}{$p_{T}>0.2$, $\rm {p} <1.6$ } &  &  \\ \cline{2-3}
\multicolumn{1}{l|}{} & \multicolumn{1}{l|}{$\Lambda,\bar{\Lambda},K^{0}_{S}$, $\Xi$, $\overline{\Xi}$} & \multicolumn{1}{l|}{$0.2<p_{T}<5.0$ } &  &  \\ \cline{2-3}
\multicolumn{1}{l|}{} & \multicolumn{1}{l|}{$\phi$} & \multicolumn{1}{l|}{$0.15<p_{T}<10.0$ } &  &  \\ \cline{2-3}
\label{tab1111}
\end{tabular}
\caption{List of hadrons with their corresponding momentum cuts used in this analysis.}
\end{table}

The directed flow is calculated by averaging the azimuthal angle ($\phi$) using the formula $\vv$ = $\langle {\rm cos}(\phi-\Psi_{RP}) \rangle$ with
respect to the reaction plane angle, $\Psi_{RP}$.

\begin{figure*}
\includegraphics[scale=0.73]{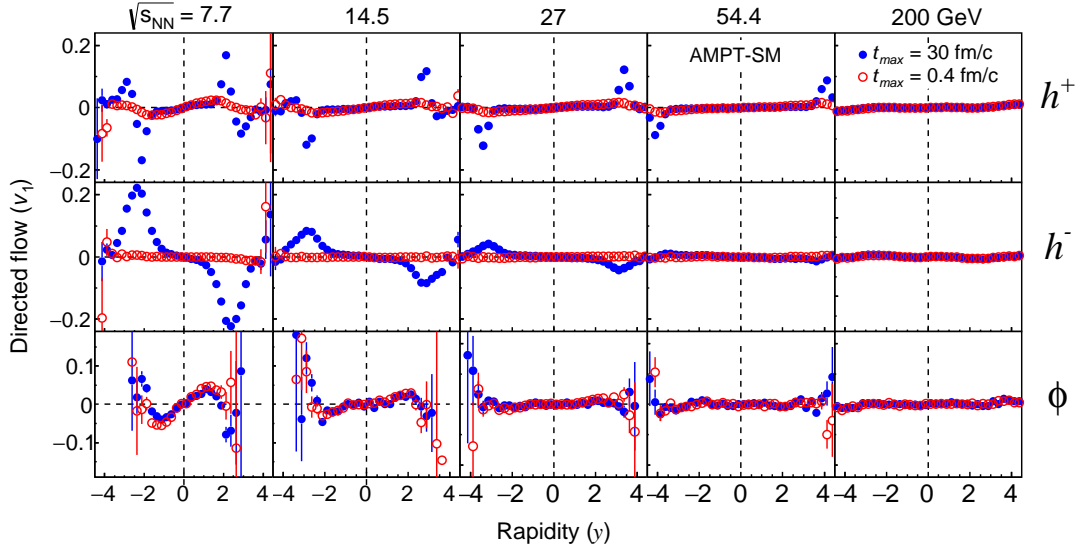}
\caption{Directed flow ($\vv$) as a function of rapidity (y) for hadron cascade time, $t_{max}$ = 30 fm/$c$ (solid marker), 0.4 fm/$c$ (open marker). Upper, middle and lower rows correspond to positively, negatively charged hadrons and $\phi$ meson, respectively in 10-40$\%$ centrality,  Au+Au collisions at $\sq$ = 7.7, 14.5, 27, 54.4 and 200 GeV using AMPT-SM.} 
\label{ntmax}
\end{figure*}

Figure~\ref{ntmax} shows the directed flow of charged hadrons and $\phi$ mesons as a function of rapidity for $t_{max}$ = 0.4 and 30 fm/$c$ in 10-40$\%$ centrality, Au+Au collisions at 7.7, 14.5, 27, 54.4 and 200 GeV. The $\vv$ of negatively charged hadrons (and positively charged hadrons at higher energies) are found to be not well developed for $t_{max}$ = 0.4 fm/$c$, because the particles could not get enough time to have hadronic interactions unlike the case of $t_{max}$ = 30 fm/$c$. However, positively charged hadrons at lower energies for $t_{max}$ = 0.4 fm/$c$ have relatively significant $\vv$ as compared to higher energies because of the dominant transported quark at low energies. It is also observed that the hadronic interaction affects $\vv$ more at higher rapidity.  The $\phi$ mesons in our $t_{max}$ = 30 fm/$c$ results represent those which have not decayed by the time of 30 fm/$c$ (i.e. $\phi$ mesons that have survived to the time of 30 fm/$c$). However, $\phi$ meson life time is relatively large and hence they represent a majority of the total $\phi$ mesons. The $\vv$ of $\phi$ meson for $t_{max}$ = 0.4 and 30 fm/$c$ are found to be similar at higher energies i.e unaffected by hadronic interactions. This is because the $\phi$ meson has a small hadronic scattering cross section and long life time ($\sim$ 42 fm/$c$), which thus leads to its decay mainly outside the fireball~\cite{pdg}. We also find that hadronic scatterings have little effect on the proton and anti-proton $\vv$ within $|y|<1.5$. 

\begin{figure*}
\includegraphics[scale=0.82]{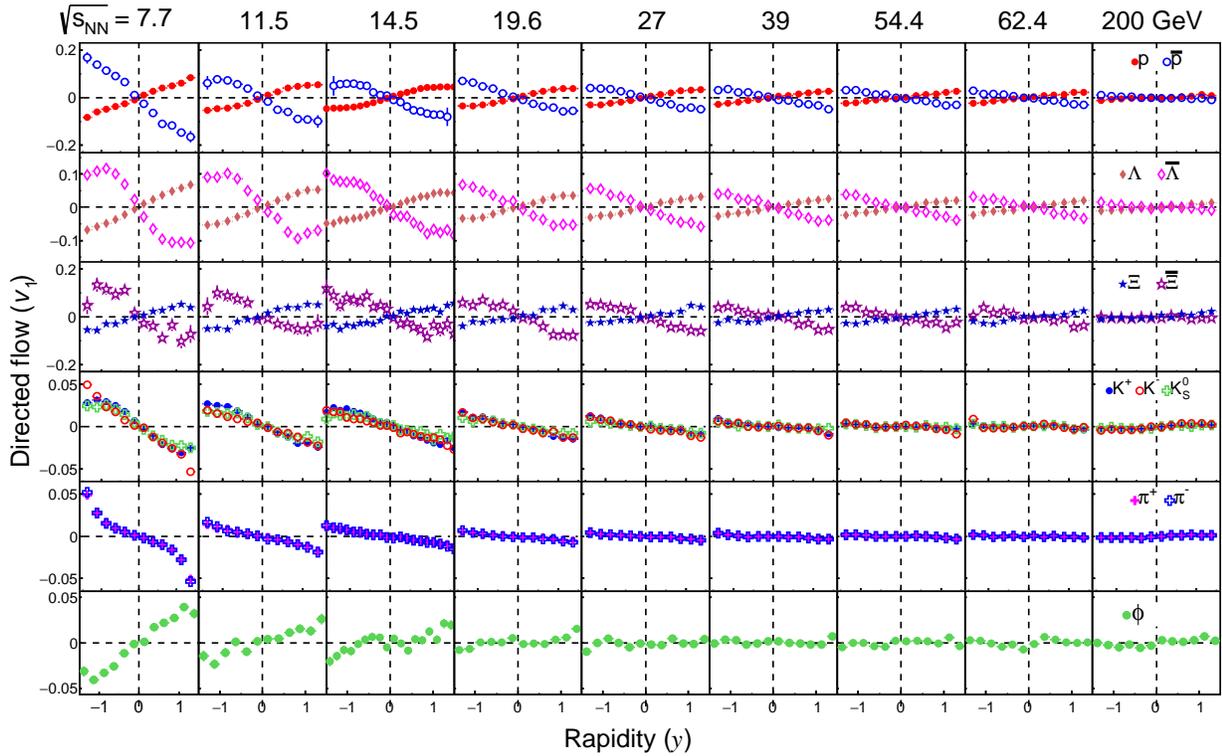}
\caption{(Color online) Directed flow ($\vv$) as a function of rapidity (y) for hadron cascade time, $t_{max}$ = 30 fm/$c$ for different identified hadrons (rows) in Au+Au collisions at $\sq$ = 7.7, 11.5, 14.5, 19.6, 27, 39, 54.4, 62.4 and 200 GeV (columns).}
\label{v1hadronAMPT}
\end{figure*}

Figure~\ref{v1hadronAMPT} shows the directed flow of various identified hadrons (corresponding rows) as function of rapidity in semi-central (10-40$\%$) Au+Au collisions at different collision beam energies (corresponding columns) using AMPT-SM, $t_{max}$ = 30 fm/$c$. The rapidity dependence of identified hadrons $\vv$ gets stronger with decreasing collision beam energy. At highest RHIC energy (200 GeV), particle and anti-particles $\vv$ values are found to be similar. The $\vv$ values of baryons and anti-baryons have opposite trend and the difference increases with decrease with energy. The mesons like $K^{\pm}$ and $K^{0}_{S}$ have similar $\vv$ values like $\pi^{+}$ and $\pi^{-}$ over the measured beam energies. $\phi$ meson $\vv$ as a function of rapidity is observed to be similar to baryons ($p$, $\Lambda$, $\Xi$), which have a strong positive slope at lower energy unlike other mesons ($K^{\pm},~K^{0}_{S},~\pi^{\pm}$). 

\begin{figure*}
\includegraphics[scale=0.84]{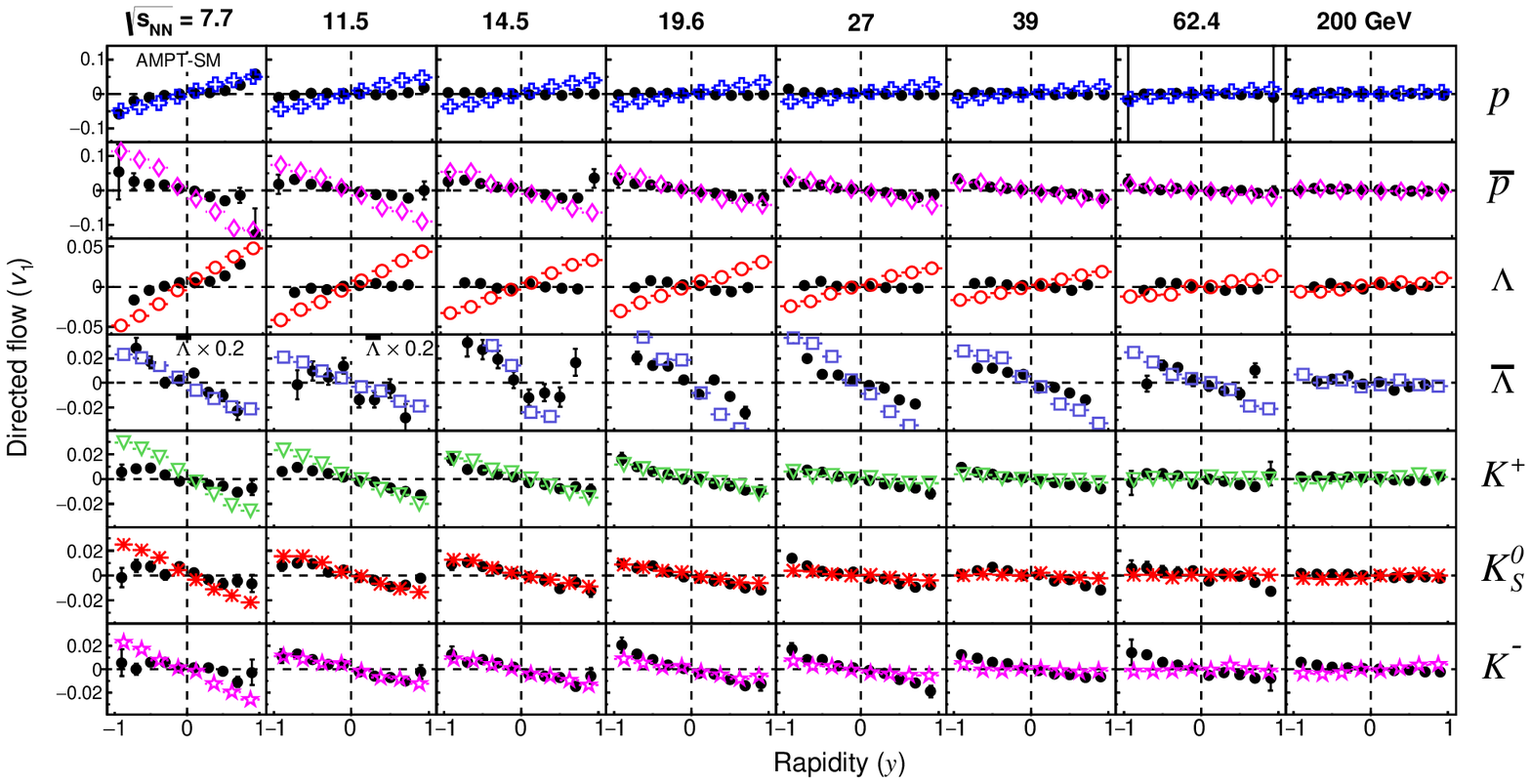}
\caption{(Color online) Directed flow ($\vv$) as a function of rapidity (y) for different identified hadrons (rows) using AMPT-SM model (hadron cascade time, $t_{max}$ = 30 fm/$c$) is compared with the experimental data (solid circle) from STAR at RHIC~\cite{starPRL} in Au+Au collisions at $\sq$ = 7.7, 11.5, 14.5, 19.6, 27, 39, 62.4 and 200 GeV (columns).}
\label{v1expAMPT}
\end{figure*}

Figure~\ref{v1expAMPT} shows the comparison of directed flow as a function of rapidity between experimental data from STAR at RHIC~\cite{starPRL} and AMPT-SM ($\sigma_{p}$ = 1.5 mb, $t_{max}$ = 30 fm/$c$) calculation for different identified hadrons at various collision energies. The AMPT-SM model better describes the experimental data of mesons as compared to the baryons and anti-baryons over the studied energy range. 

The strength of directed flow signal at mid-rapidity is usually characterized by the linear term, F, in the equation $v_{1}(y)$ = $Fy+F_{3}y^{3}$~\cite{starPRL} or by the slope ($F^{'}$) parameter of the fit function $v_{1}(y)$ = $F^{'}y+C$~\cite{protonv1}. Here, the slope parameter $F^{'}$ is denoted as $\dv$.  By using the cubic fit function one can reduce sensitivity to the rapidity range in which the fitting is performed. However, in order to have a better comparison we have used the linear fit function similar to the experimental STAR result~\cite{starPRL}. The fitting range for various hadron species are $|y|$ $<$ 0.8 for all measured particles except for $\phi$ meson which is fitted in the rapidity range, $|y|$ $<$ 0.6.    

\begin{figure}
\includegraphics[scale=0.35]{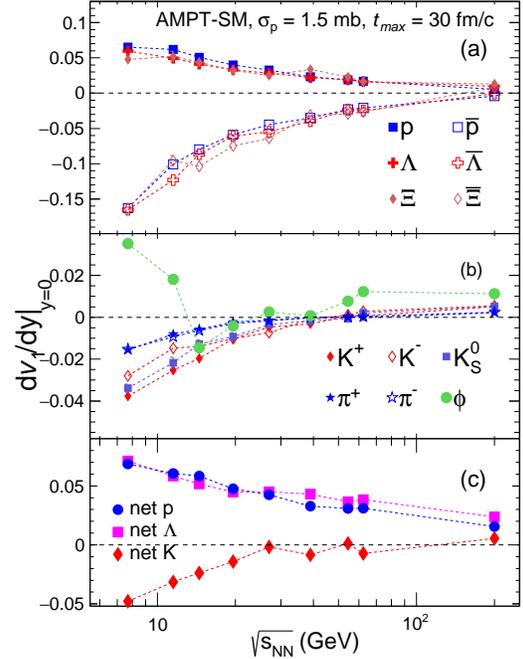}
\caption{(Color online) The slope ($\dv$) of baryons, mesons and net-$p$, net-$\Lambda$, net-$K$ are shown in upper, middle and lower panels as a function of beam energy for 10-40$\%$ centrality, respectively. The dotted lines are smooth curves drawn here to guide the eye.}
\label{slope_hadrons}
\end{figure}

The collision beam energy dependence of the directed flow slope, $\dv$, for baryons ($p$, $\overline{p}$, $\Lambda$, $\overline{\Lambda}$, $\Xi$, $\overline{\Xi}$) and mesons ($\pi^{-}$, $\pi^{+}$, $K^{+}$, $K^{-}$, $K^{0}_{S}$, $\phi$) are shown in Fig.~\ref{slope_hadrons} (a) and Fig.~\ref{slope_hadrons} (b), respectively. The $\dv$ of measured baryons such as $p$, $\Lambda$, and $\Xi$ are found to have similar value and their anti-particles $\bar{p}$, $\bar{\Lambda}$, and $\overline{\Xi}$ have also similar slope within the uncertainty over the measured energy range. All the measured baryons have positive d$v_1$/dy where as their anti-particles have negative slope values. In AMPT-SM, the sign change of baryons' ($p$ and $\Lambda$) slope is not observed unlike observed in STAR experiment~\cite{starPRL}.  The $\dv$ of $\pi^{-}$ and $\pi^{-}$ are similar; $K^{+}$ and  $K^{-}$ values are also similar except for lower energies ($<$ 19.6 GeV) and their average value corresponds to slope of $K^{0}_{S}$ meson. All the mesons except $\phi$ resonance have negative $\dv$ below 39 GeV collision energy like the corresponding STAR results~\cite{starPRL}. Overall magnitude of baryons and anti-baryons $\dv$ are larger than the mesons.  

\begin{figure}
\includegraphics[scale=0.35]{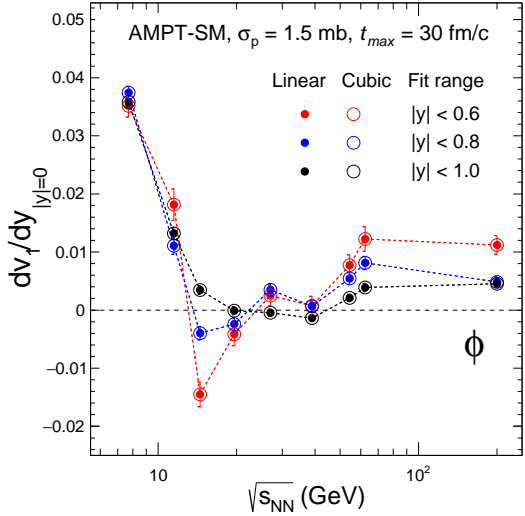}
\caption{(Color online) Beam energy dependence of $\phi$-meson slope parameter obtained using different fitting ranges and fit functions for 10-40$\%$ centrality. The dotted lines are smooth curves drawn here to guide the eye.}
\label{slope_phi_fit}
\end{figure}

Figure~\ref{slope_phi_fit} shows the $\phi$ meson slope calculated by using different fitting ranges for both linear and cubic function in Au+Au collisions from $\sq$ = 7.7-200 GeV. The $\dv$ shows the sign change in between 11.5 to 27 GeV for the fitting range $|y|$ $<$ 0.6 and $|y|$ $<$ 0.8. The fitting range  in the STAR measurement is $|y|$ $<$ 0.6~\cite{starPRL}. When the fitting range increases, the magnitude of negative slope decreases and for $|y|$ $<$ 1.0 the $\dv$ becomes positive within uncertainty for all measured energies. There is an hint of slope change as observed in STAR~\cite{starPRL} although the statistical significance is poor. The slope change of $\phi$ meson might be due to short range fitting ($|y|$ $<$ 0.6) of $\vv$ as a function of rapidity. There is no difference between the $\phi$ meson slope calculated using linear and cubic function even though different fitting ranges are considered.  One can also observed that there is a sharp increase in the $\phi$ meson slope with decrease in energy ($<$ 11 GeV) which is similar to the STAR experimental results at RHIC~\cite{starPRL}. 

The energy dependence of proton $\dv$ receives contribution mainly in two ways (i) $\vv$ of transported protons from the initial colliding beam rapidity toward the mid-rapidity and (ii) $\vv$ of protons from pair (particle and anti-particle) production near mid-rapidity. The importance of pair production increases with increase in colliding energy. The "net particle" is a measure of excess particles yield over its anti-particles. It is used to disentangle the transported quarks relative to that of produced in the collisions by using Eq.~\ref{net}. 

\begin{equation}
[v_{1}(y)]_{p} = r(y)[v_{1}(y)]_{\bar{p}} + [1-r(y)][v_{1}(y)]_{{\rm net}-p},
\label{net}
\end{equation}

where $r(y)$ is the rapidity dependence of anti-proton to proton ratio at each beam energy. The formulae for net-$K$ and net-$\Lambda$ are defined in the similar way as Eq.~\ref{net}.  Anti-proton $\vv$ has been proposed as proxy of produced proton $\vv$ in the Ref~\cite{protonv1} and net-$p$ slope is also used to distinguish the transported baryonic matter and hydrodynamic effect~\cite{protonv1,starPRL}. There are also model calculation which suggests that the transported quarks ($u$ and $d$ from initial colliding nuclei) contribute more towards the formation of hadrons like $p$, $\Lambda$ and $K^{+}$~\cite{transport1}. Figure~\ref{slope_hadrons} (c) shows the net-$p$, net-$\Lambda$ and net-$K$ $\dv$ as a function of beam energy for mid-central (10-40$\%$) Au+Au collisions from $\sq$ = 7.7-200 GeV. The net-$p$ and net-$\Lambda$ have positive and similar $\dv$ unlike the net-$K$ over the measured energy range.    

In this analysis, there are several (12) hadrons which allow us to have a comprehensive study of constituent quark $\vv$. The assumption like $\vv$ is developed in pre-hadronic stage, each type of quark has different directed flow and that hadrons are formed via quark coalescence can be tested here. The coalescence sum rule suggests that at smaller azimuthal anisotropy coefficient ($v_{n}$), the detected hadron's $v_{n}$ is sum of their constituent quark's $v_{n}$~\cite{starPRL}. The popular example of NCQ scaling observed at RHIC and LHC are followed from the coalescence sum rule~\cite{ncqSTAR,ncqPHENIX,ncqALICE}. 

\begin{figure}
\includegraphics[scale=0.37]{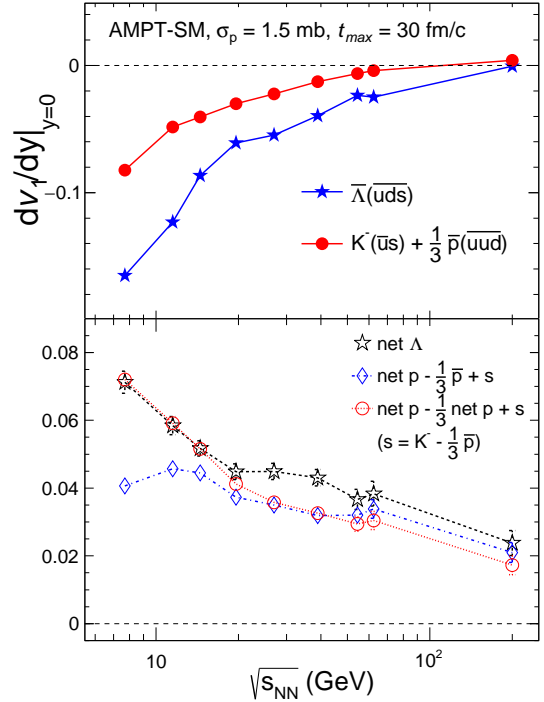}
\caption{(Color online) Upper panel shows the comparison of $\bar{\Lambda}(\overline{uds})$ and sum rule test for produced quark ($K^{-}(\bar{u}s)+\frac{1}{3}\bar{p}(\overline{uud})$) slope as a function of beam energy. Lower panel shows another set of sum rule test using net-$\Lambda$ and net-$p$ for 10-40$\%$ centrality in Au+Au collisions using AMPT-SM. The solid and dotted lines are smooth curves drawn here to guide the eye.}
\label{slope_Nethad}
\end{figure}

\begin{figure*} 
\includegraphics[scale=0.83]{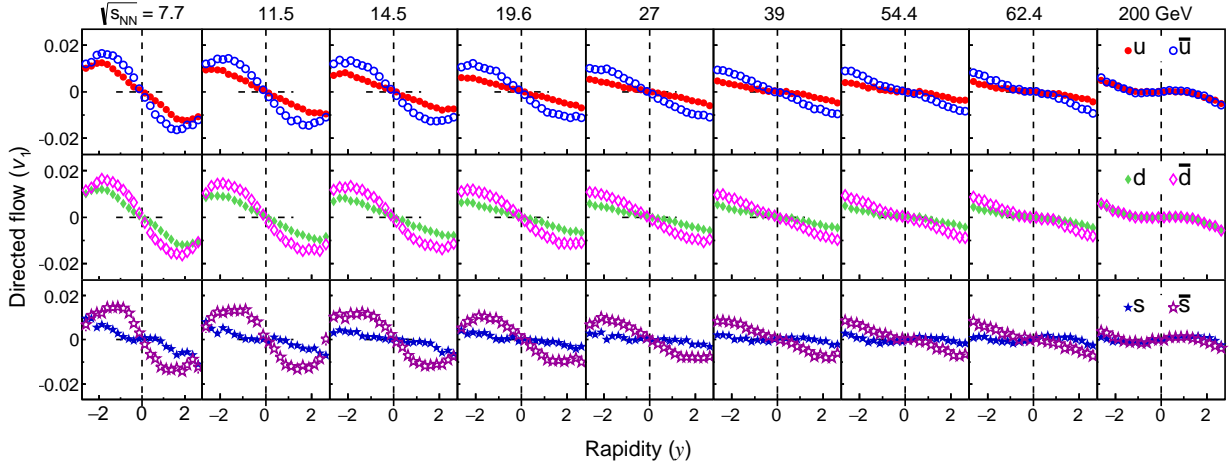}
\caption{(Color online) Directed flow ($\vv$) as a function of rapidity (y)  for different quark (anti-quark) solid marker (open marker) are shown in corresponding rows for Au+Au collisions at $\sq$ = 7.7, 11.5, 14.5, 19.6, 27, 39, 54.4, 62.4 and 200 GeV (columns) using AMPT-SM.}
\label{quarkv1}
\end{figure*} 

\begin{figure}[H]
\begin{center}
\includegraphics[scale=0.4]{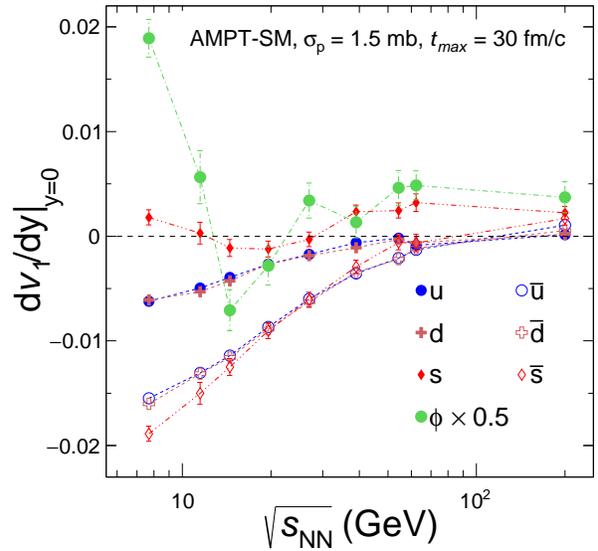}
\caption{(Color online) Slope ($\dv$) as a function of $\sq$ for quark, anti-quark and $\phi$ meson using AMPT-SM for parton-parton cross section ($\sigma_{p}$) of 1.5 mb. The $\phi$ meson slope is divided with corresponding number of constituent quarks i.e 2.}
\label{quarkSlope}
\end{center}
\end{figure}

Figure~\ref{slope_Nethad} (upper panel) shows the comparison of $\bar{\Lambda}(\overline{uds})$ and $K^{-}(\overline{u}s)+\frac{1}{3}\overline{p}(\overline{uud})$ slope as a function of beam energy for 10-40$\%$ centrality in Au+Au collisions from 7.7-200 GeV. The example stated here is the most suitable to test coalescence sum rule because both $\overline{\Lambda}(\overline{uds})$ and $\overline{p}(\overline{uud})$ are produced unlike the $u$ and $d$ quarks which could be either produced or transported. However, by comparing these two cases we have assumed that $s$ and $\overline{s}$ have same flow. The scale factor $\frac{1}{3}$ is due to the assumption that $\overline{u}$ and $\overline{d}$ have the same $\vv$. But we found that except for highest energy both of them are found to have different slope indicating violation of these assumptions. The $\dv$ of $s$ and $\bar{s}$ are different except for highest RHIC energy as shown in the Fig~\ref{quarkSlope} for AMPT-SM, $\sigma_{p}$ = 1.5 mb. As per the assumption, one can observe that $\overline{u}$ and $\overline{d}$ have similar slope as shown in the Fig.~\ref{quarkSlope}. The STAR measurement at RHIC also found that the slope of $\bar{\Lambda}(\overline{uds})$ and scaled $\overline{p}(\overline{uud})$ have different slope at lower energies~\cite{starPRL} and this might be due to the assumption that $s$ and $\bar{s}$ have similar $\vv$ over all measured energy ranges which may not be valid for lower energies. 

Figure~\ref{slope_Nethad} (lower panel) shows the first case of coalescence sum rule involving $u$ and $d$ quarks which are either transported or produced and hence it is cumbersome to distinguish them in general. However, one can naively expect that at lower beam energy, $u$ and $d$ quarks are mostly transported whereas these quarks are largely produced at high colliding beam energy. In this figure, two different coalescence sum rule scenarios are compared with the net-$\Lambda$ (open star). First case is the net-$p$ minus $u$ plus $s$, where $u$ and $s$ quarks are obtained from $\bar{p}/3$ and $K^{-}(\bar{u}s) -\frac{1}{3}\bar{p}(\overline{uud})$, respectively as represented by blue diamond symbol. Here, the produced $u$ quark in net-$p$ is replaced by $s$ quark. However, we do not have the corresponding straight forward expression for representing the transported u and d quarks. The sum rule is found to be in a good agreement with net-$\Lambda$ above 39 GeV and start deviating for lower energies. This observation suggests that the fraction of transported quarks in the constituent quarks assembly of net-$p$ increases with decrease in collision beam energy, which imply that the assumption of produced $u$ quark is removed by keeping the term (net-$p$ - $\frac{1}{3}\bar{p}$) also starts deviating. The observation of getting the transported quark dominance at lower energy ($\leq$ 39 GeV) in 10-40$\%$ centrality Au+Au collisions using AMPT-SM is qualitatively similar to that of observed in STAR experiment at RHIC~\cite{starPRL}. 

The second case of coalescence sum rule i.e ($\frac{2}{3}$net-$p$ + s) is also shown in (red open circle marker) Fig.~\ref{slope_Nethad} (lower panel). In this sum rule, it is assumed that in the limit of low beam energy, the constituent quarks of net protons are dominated by transported quarks, and $s$ quark replaces one of the transported quarks. This assumption starts showing disagreement for beam energy greater than 19.6 GeV i.e. the disagreement between black and red markers. 

Figure~\ref{quarkv1} shows the directed flow of partons ($u$, $\bar{u}$, $d$, $\bar{d}$, $s$ and $\bar{s}$) systematic evolution in Au+Au collisions from low to high energy ($\sq$ = 7.7 to 200 GeV). All the anti-quarks are produced unlike the $u$ and $d$ quarks which might be either transported or produced depending on the collisions beam energy. So, at highest RHIC energy both the quarks and anti-quarks have same $\vv$ as these are expected to be mostly produced. However, with decrease in beam energy the $\vv$ difference between them increases and anti-quarks shows larger directed flow than quarks. $\vv$ of quarks forming primordial proton have the opposite sign compared with $\vv$ of all quarks, and further study is needed to understand why this is the case for the quark coalescence in AMPT. Furthermore, the $\vv$ slope of quarks coalescing to primordial proton and the corresponding proton have similar slope. However, the final proton (Fig.~\ref{v1hadronAMPT}) which includes decay contribution and hadronic interaction have similar positive slope like primordial proton but different in magnitude.



Directed flow slope parameter of quarks, anti-quarks and $\phi$ meson as a function of beam energy is shown in Fig.~\ref{quarkSlope}. The slope of $u$ and $d$ quarks is found to be similar and decreases with increase in beam energy unlike the $s$ quark. All the light anti-quarks ($\bar{u}$, $\bar{d}$ and $\bar{s}$) have more negative slope than their corresponding quarks. However, there is a clear deviation of $\bar{s}$ slope from the trend of $s$ quark except at the highest energy. The $\phi$ meson slope does not scale with the $s$ and $\overline{s}$ quarks' slope. 

\section{Summary and outlook}   
\label{s-End}
A comprehensive study of rapidity-odd component directed flow for charged and identified hadrons in Au+Au collisions (10-40$\%$ centrality) for a range of collision beam energies using an improved coalescence AMPT-SM model has been discussed. The coalescence sum rule or commonly known as NCQ scaling is tested using the directed flow measurement of identified hadrons. The analysis performed here are summarized in the following. 

The effect of hadronic interaction on $\vv$ of charged hadrons and $\phi$ meson are reported. The $\vv$ of charged hadrons are found to be not well developed for $t_{max}$ = 0.4 fm/$c$, because the particles could not get enough time to have hadronic interactions unlike the case for $t_{max}$ = 30 fm/$c$, except for positively charged hadrons at lower energies where the transported quark effect is more dominant. However, the $\phi$-meson $\vv$ is found to be unaffected by hadronic interaction because of it's small hadronic cross section and also it decays outside the fireball (life time $\sim$ 42 fm/$c$). The double sign change of $\phi$ meson slope in between 11.5 to 27 GeV is observed. This sign change is also found to be an artifact of small fitting ranges while extracting the $\vv$ slope. The sign change goes away making positive slope for all measured energies when the linear or cubic function is fitted in a larger rapidity range ($|y|$ $<$ 1.0). This observation emphasizes the crucial importance of fitting range while extracting the slope parameter (d$\vv$/dy) in the real data measurement. Prediction for directed flow as function of rapidity and slope parameter of various identified hadrons in semi-central (10-40$\%$) Au+Au collisions at $\sq$ = 54.4 GeV are reported. The $\vv$ calculation of $\Xi$ and $\overline{\Xi}$ baryons is predicted for a range of energy and the values are found to be similar to protons and $\Lambda$ baryons. The $\vv$ results at higher rapidity range are also shown here, which cover the Event Plane Detector (EPD) pesudo-rapidity ($\eta$) range installed in STAR detector at RHIC for BES-II program.

We find that light quarks such as $u$ and $d$ have similar slope and it decreases with increase in beam energy unlike the $s$ quark. The anti-quarks ($\bar{u}$, $\bar{d}$ and $\bar{s}$) have more steeper negative slope than corresponding light quarks and are similar for the measured beam energy range. The $s$ and $\overline{s}$ quarks have different $\vv$ except for the highest energy. There is a clear indication that $\overline{s}$ quark slope start deviating from the trend of $s$ quark with the decrease in energy. The measured baryons ($p$, $\Lambda$ and $\Xi$) have similar positive slope and increases with decrease in beam energy unlike their corresponding anti-particles. The AMPT-SM model shows no sign change for $p$ and $\Lambda$ slope unlike that observed in STAR experiment at RHIC~\cite{starPRL}. The slopes of $\pi^{+}$, $\pi^{-}$, $K^{+}$, $K^{-}$ and $K^{0}_{S}$ mesons are positive at the highest RHIC energy then start decreasing and becomes negative with decrease in beam energy. The slope of $K^{0}_{S}$ is approximately average of $K^{+}$ and $K^{-}$ slope as observed in STAR at RHIC~\cite{starPRL}.

The test of coalescence sum rule using produced quarks are done by comparing the slope of $\bar{\Lambda}(\overline{uds})$ and $K^{-}(\bar{u}d)+\frac{1}{3}\bar{p}(\overline{uud})$. These are found to have different slope and the departure increases with decrease in energy which might be due to break-down of the assumption that $s$ and $\bar{s}$ have same flow over the measured energy range. The slope of net-$p$ and net-$\Lambda$ are similar over the measured energy range. The sum rule (net-$p$ - $\frac{1}{3}$$\bar{p}$ + $s$) and net-$\Lambda$ are found to be similar for energy higher than 39 GeV. The deviation at lower energy might be an indication that the assumption of produced $u$ quarks effect can be removed by keeping the term $\frac{p}{3}$. This assumption does not hold at lower energies, which is similar to the observation in STAR at RHIC~\cite{starPRL}. The sum rule ($\frac{2}{3}$net-$p$ + $s$) and net-$\Lambda$ values starts deviating at energy higher than 19 GeV. This sum rule assumes that at lower energy the transported quarks dominates and one of the transported quark of net-$p$ is replaced by $s$ quark. Hence, this approximation breaks down in the limit of high beam energy which is qualitatively similar to the observation in STAR at RHIC~\cite{starPRL}.  


\section{ACKNOWLEDGMENTS}   
 This work is supported in part by the National Natural Science Foundation of China under Grants No. 11890711 and self-determined research funds of CCNU from the colleges basic research and operation of MOE under Grant No. CCNU18TS031.



\end{document}